\documentclass[%
 aip,
 pop
 amsmath,amssymb,
 reprint,%
]{revtex4-1}

\usepackage{graphicx}
\usepackage{dcolumn}
\usepackage{bm}

\newcommand{\sss}{\scriptscriptstyle}
\newcommand {\be}{\begin{equation}} 
\newcommand{\ee}{\end{equation}}    

\def\vti{v_{{\sss T}i}}
\def\vte{v_{{\sss T}e}}
\def\vtj{v_{{\sss T}j}}
\def\vta{v_{{\sss T}a}}
\def\vtb{v_{{\sss T}b}}
\def\vt{v_{{\sss T}}}

\begin{document}
\preprint{AIP/123-QED}

\title[]{A new low-frequency backward mode in inhomogeneous plasmas }

\author{J. Vranjes}
  \email{jvranjes@yahoo.com}
 \affiliation{
Institute of Physics Belgrade, Pregrevica 118, 11080 Zemun, Serbia\\
}%

\date{\today}

\begin{abstract}
When an electromagnetic transverse wave propagates through an inhomogeneous plasma so that its electric field has a component in the direction of the background density gradient, there appears a disbalance of charge in every plasma layer, caused by the density gradient. Due to this some additional longitudinal electric field component appears in the direction of the wave vector. This longitudinal field may couple with the usual electrostatic longitudinal perturbations like the ion acoustic, electron Langmuir, and ion plasma waves. As a result, these standard electrostatic waves are modified and in addition to this a completely new low-frequency mode appears. Some basic features of the coupling and modification of the ion acoustic wave, and properties of the new mode are discussed here, in ordinary electron-ion and in pair plasmas.

%
\end{abstract}

\maketitle



\section{\label{s1} Introduction }
%

In ordinary electron-ion plasmas, both transverse electromagnetic (TEM) and longitudinal electrostatic (LES) Langmuir wave have the same  cut-off at the electron plasma frequency $\omega_{pe}$. For very small wave-numbers  the frequencies of both modes are close to each other. The two modes are physically very different and within linear theory they are typically not coupled. On the other hand, ion modes [like ion acoustic (IA), and ion plasma (IP) modes\citep{ses, duc, vk}] are  well separated from the two mentioned modes and it is believed that there is no linear coupling either. However, the situation may be quite different in the presence of a density gradient in the direction perpendicular to the direction of propagation of the TEM wave. In such a case the LES modes (IA, IP, and Langmuir)
 become coupled with the TEM mode even within linear theory. The coupling is more profound for the IA mode  \citep{vaa}, implying that these electrostatic modes may have some  electromagnetic features.
  The coupling with the IA mode has been studied in detail in Ref.~4 for both cold and hot ions, collisional and collision-less, isothermal and adiabatic.
  It was shown that in a part of spectrum, for small wave-numbers $k$,  the IA mode may become backward in the sense that $\partial \omega/\partial k<0$, it gets some cut-off caused by the density gradient, and in this same domain it is coupled with TEM wave.

  However, in the previous work  \citep{vaa} it was not realized that the general dispersion equation, which describes coupling between TEM and LES waves in the presence of a density gradient, allows for some additional peculiar low frequency hybrid (LFH) mode  in the range below  the IA wave frequency $k c_s$ and below the ion thermal mode $k \vti$, where $c_s, \vti$ are the sound and ion thermal speeds, respectively. In {\bf the  large} part of the spectrum this  LFH mode is backward, $\partial \omega/\partial k<0$, and it appears only in the presence of a TEM wave propagating through an inhomogeneous environment.  The mode is the result of linear coupling between TEM and LES modes. At frequencies close to $k c_s$ there is an exchange of identities of this new mode and the IA mode; the latter becomes backward above $k c_s$ for small wave-numbers\citep{vaa}, while the LFH mode  follows the $k c_s$ line (but remaining below it) for $k\rightarrow 0$. Some basic features of this new low-frequency mode, and its coupling with the IA mode are presented in this work.
\section{\label{s2}Plasma without magnetic field}
We start with a static plasma containing two general species $a$ and $b$, which thus may include some ion-electron or pair (pair-ion, electron-positron) plasma, and  we assume small isothermal electromagnetic perturbations that propagate in $z$-direction. Note that much more general cases were studied in our recent work \citep{vaa}, for collisional plasma with hot ions, with the Landau damping effect, and together with the energy equation. In the present case we take a simple model in order to see some  basic features of the new low-frequency mode, presented here for the first time.   Linear  perturbations imply the momentum equation for the general species $j$:
\be
m_j n_{j0}\frac{\partial \vec v_{j1}}{\partial t} = q_j n_{j0} \vec E_1-\kappa T_{j0}\nabla n_{j1} - \kappa n_{j1} \nabla T_{j0}. \label{e1}
\ee
Here, indices $0, 1$ describe the equilibrium and perturbed quantities, respectively, for the two species $j=a, b$,  where $q_a=e$, $q_b=-e$, $n_{j0}=n_0$,  and $T_j\equiv T_{j0}$.
We shall assume small  equilibrium gradients of the temperature and density to be in $x$-direction, and  with the characteristic inhomogeneity length far exceeding the wave-length, so that we apply the usual local approximation analysis. We allow for the presence of both longitudinal (electrostatic) and transverse (electromagnetic)  perturbations  propagating in the $z$-direction,  $\sim -i \omega t + i k z$. The  speed due to both of these perturbations
\be
\vec v_{j1}= \frac{i q_j \vec E_1}{m_j \omega} - \frac{i \vtj^2}{\omega} \frac{\nabla n_{j1}}{n_0} - \frac{i \vtj^2}{\omega} \frac{n_{j1}}{n_0}
\frac{\nabla T_j}{T_j}, \label{e3}
\ee
is used in the continuity equation which becomes
\be
- i \omega \frac{n_{j1}}{n_0} + \nabla\cdot \vec v_{j1} + \vec v_{j1} \frac{\nabla n_0}{n_0}=0. \label{e4}
\ee
In what follows we use the fact that $\vec k\bot \nabla n_0$,  $\vec k\bot\nabla T_j$ and without any approximation from Eq.~(\ref{e4}) we have
\be
n_{j1}=\frac{q_j n_0}{m_j \omega_{j}^2} \nabla\cdot\vec E_1 + \frac{q_j  }{m_j \omega_{j}^2}\vec E_1\cdot \nabla n_0,
 \label{e5}
\ee
\[
\omega_{j}^2=\omega^2- k^2 \vtj^2 + \frac{\vtj^2 T_{j}''}{T_{j}}.
\]
Here, $n_0'$, $T_{j}''$ are  the first and second derivatives  of $n_0, T_{j}$ in the $x$-direction.  The electric field in principle has both longitudinal and transverse components; the one  in the term with the density gradient is due to transverse plane-polarized electromagnetic perturbations. It produces  the term with $E_{1x} n_0'$ which contributes to density perturbations and it is therefore responsible for the coupling between longitudinal and transverse oscillations (this is described in more details below, see Fig.~\ref{mech}). Clearly, for the purpose of the present work, this transverse $\vec E_1$   can be in any direction perpendicular to $z$, except strictly perpendicular to the density gradient (i.e., in the $y$ direction). So for simplicity, in what follows we shall  assume it to be in the plane of the density gradient (the $x$-plane).

We need $\nabla n_{j1}$ in Eq.~(\ref{e3}),  and from Eq.~(\ref{e5}) the result is
\[
\nabla n_{j1}=\frac{q_j n_0}{m_j \omega_j^2} \nabla (\nabla\cdot \vec E_1) + \frac{q_j  \nabla\cdot \vec E_1}{m_j \omega_j^2} \nabla n_0
+ \frac{q_j n_0'}{m_j \omega_j^2} \nabla E_{1x}
\]
\[
+  \frac{q_j E_{1x}}{m_j \omega_j^2} \nabla n_0' + \frac{ k^2\vtj^2}{\omega_j^4} \frac{q_j E_{1x} n_0'}{m_j} \frac{\nabla T_{j}}{T_{j}}
\]
\be
+  \frac{ k^2\vtj^2}{\omega_j^4}\frac{q_jn_0\nabla\cdot\vec E_1}{m}
  \frac{\nabla T_{j}}{T_{j}}. \label{e6}
  \ee
  In this expression only one term containing the third derivative $\vtj^2\nabla T_{j}''/(\omega_j^4 T_{j})$ has been neglected.
Eqs.~(\ref{e5},~\ref{e6}) are used  in Eq.~(\ref{e3}) which becomes
\[
\vec v_{j1}=\frac{i q_j}{m_j \omega} \vec E_1 - \frac{i q_j\vtj^2}{m_j \omega\omega_j^2} \nabla (\nabla\cdot\vec E_1)
 \]
 \[
 - \frac{i q_j\vtj^2 \nabla\cdot\vec E_1}{m_j \omega\omega_j^2} \left(\frac{\nabla n_0}{n_0} + \frac{\nabla T_{j}}{T_{j}}\right)  -  \frac{i q_j\vtj^2 E_{1x}}{m_j \omega\omega_j^2} \frac{n_0'}{n_0} \frac{\nabla T_{j}}{T_{j}}
\]
\[
-  \frac{i q_j\vtj^2}{m_j \omega\omega_j^2} \left(E_{1x} \frac{\nabla n_0'}{n_0} + \nabla E_{1x} \frac{n_0'}{n_0}\right)
\]
\be
- \frac{i q_jk^2 \vtj^4}{m_j \omega\omega_j^4} \frac{\nabla T_{j}}{T_{j}} \left( E_{1x} \frac{n_0'}{n_0} +\nabla\cdot\vec E_1\right).
\label{e7}
\ee
 We may omit the  third order small terms, and the velocity becomes
\[
\vec v_{j1}=\frac{i q_j}{m_j \omega} \vec E_1 - \frac{i q_j\vtj^2}{m_j \omega\omega_j^2} \nabla (\nabla\cdot\vec E_1)
 \]
 \[
 - \frac{i q_j\vtj^2 \nabla\cdot\vec E_1}{m_j \omega\omega_j^2} \left(\frac{\nabla n_0}{n_0} + \frac{\nabla T_{j}}{T_{j}}\right)
\]
\be
-  \frac{i q_j\vtj^2}{m_j \omega\omega_j^2} \frac{n_0'}{n_0} \nabla E_{1x}  - \frac{i q_jk^2 \vtj^4}{m_j \omega\omega_j^4} \frac{\nabla T_{j}}{T_{j}}
 \nabla\cdot\vec E_1.
\label{e8}
\ee
Here and further in the text  the second derivative of temperature in $\omega_{j}$ is neglected,  in accordance with the used local approximation.
 Eq.~(\ref{e8}) is used in the general wave equation
\be
c^2k^2\vec E_1- c^2 \vec k (\vec k\cdot \vec E_1) - \omega^2 \vec E_1 - \frac{i \omega \vec j_1}{\varepsilon_0}=0,
\label{e9}
\ee
\[
\quad \vec j_1= n_0\left(q_a \vec v_{a1}-
q_b \vec v_{b1}\right).
\]
This yields the following wave equation expressed through the perturbed electric field only:
\[
c^2 k^2 \vec E_1- c^2 \vec k \left(\vec k\cdot \vec E_1\right) - \omega^2 \vec E_1 +  \left(\omega_{pa}^2 + \omega_{pb}^2\right) \vec E_1
\]
\[
-  \frac{n_0'}{n_0} \left(\frac{\omega_{pa}^2 \vta^2}{\omega_a^2}
+ \frac{\omega_{pb}^2 \vtb^2}{\omega_b^2}\right)\nabla E_{1x}
\]
\[
- k^2 \nabla\cdot \vec E_1 \left(\frac{\omega_{pa}^2 \vta^4}{\omega_a^4} \frac{\nabla T_a}{T_a} + \frac{\omega_{pb}^2 \vtb^4}{\omega_b^4}\frac{\nabla T_a}{T_a} \right)
\]
\[
-  \left(\frac{\omega_{pa}^2 \vta^2}{\omega_a^2} + \frac{\omega_{pb}^2 \vtb^2}{\omega_b^2}\right) \nabla \left(\nabla\cdot \vec E_1\right)
\]
\[
- \nabla\cdot \vec E_1 \left[\frac{\omega_{pa}^2 \vta^2}{\omega_a^2} \left(\frac{\nabla n_0}{n_0} + \frac{\nabla T_a}{T_a}\right)
 \right.
 \]
 \be
 \left.
 + \frac{\omega_{pb}^2 \vtb^2}{\omega_b^2}
\left(\frac{\nabla n_0}{n_0} + \frac{\nabla T_b}{T_b}\right)
\right]  =0. \label{e11}
\ee
Here, $\omega_{pj}^2= e^2 n_0/(\varepsilon_0 m_j)$ and the electric field includes both transverse and longitudinal components.
\subsection{\label{m} Mechanism of transverse-longitudinal electric field coupling }

\begin{figure}[!htb]
   \centering
  \includegraphics[height=6.5cm,bb=16 14 257 218,clip=]{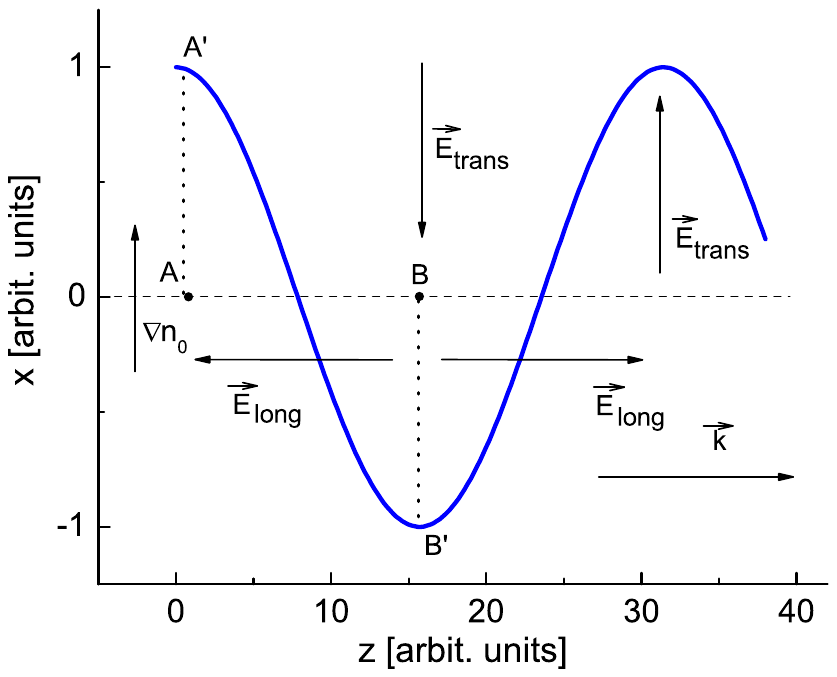}
      \caption{Origin of a longitudinal electric field in case  of a transverse electromagnetic wave propagating through an  inhomogeneous plasma.  }   \label{mech}
       \end{figure}

 The mechanism of the coupling can be understood\citep{vaa} from Fig.~\ref{mech} where we have a  density gradient in $x$-direction and an EM wave propagating along the $z$-axis, $\vec k=k\vec e_z$. Due to the electric field $\vec E_{trans}=E_{x1} \vec e_x$, plasma particles with opposite charges move in opposite directions. The physics is essentially the same for both electron-ion and pair-ion plasma, although in the former case the displacement of particles is more effective for electrons of course.  In case of particles of the same mass, displacements of positive particles, that were initially in an arbitrary layer (which we denote as $x=0$) with  density $n_0(0)$, is represented by the sinusoidal line. Two arbitrary points $A(0, z_1)$ and $B(0, z_2)$ at two different positions in $z$-direction  are displaced  to $A'(x_1, z_1)$ and $B'(x_2, z_2)$.  As a result, due to background density gradient, the amount of positively charged particles at $A(0, z_1)$ and $B(0, z_2)$ will no longer be the same; in the point $A(0, z_1)$ it is reduced (they are replaced by the particles of the same species which come from some other point in $x$ direction with lower density), and in the point $B(0, z_2)$ it is increased. In the same time, because of opposite motion of negatively charged particles, the amount of negative charges at $A(0, z_1)$ will be increased (they are displaced and moved from the area with a higher density) and at $B(0, z_2)$ decreased. Something similar happens at every point in the $z$ direction, and for every layer along the $x$ axis.  This means that there will be a difference of charge in the chosen arbitrary points, and this further implies that there will be  an additional electric field $E_{1z}$ in the $z$ direction, as indicated in Fig.~\ref{mech}.
\subsection{\label{s3} The case of equilibrium with opposite density and temperature  gradients}
 So far nothing is assumed about a possible  relation between the temperature and density gradients in the equilibrium. One possible and physically plausible  scenario  may include a plasma with a  balance of the two gradients in the equilibrium:
 \be
\frac{\nabla T_j}{T_j}= -\frac{\nabla n_0}{n_0}. \label{e2}
\ee
Note that this also allows for different temperatures of the two species $T_b=\alpha T_a$. With this, the last two terms in Eq.~(\ref{e11}) vanish and we consequently have the following wave equation:
\[
c^2 k^2 \vec E_1- c^2 \vec k \left(\vec k\cdot \vec E_1\right) - \omega^2 \vec E_1 + \left(\omega_{pa}^2 + \omega_{pb}^2\right) \vec E_1
\]
\[
 - \frac{\nabla E_{1x}}{L_n} \left(\frac{\omega_{pa}^2 \vta^2}{\omega_a^2} + \frac{\omega_{pb}^2 \vtb^2}{\omega_b^2}\right)
  \]
  \[
  +\frac{ k^2\vec e_x}{L_n}  \left(\frac{\omega_{pa}^2\vta^4}{\omega_a^4}  + \frac{\omega_{pb}^2\vtb^4}{\omega_b^4}\!\right)\!\nabla\cdot \vec E_1
\]
\be
-  \left(\frac{\omega_{pa}^2\vta^2}{\omega_a^2} + \frac{\omega_{pb}^2\vtb^2}{\omega_b^2}\right) \nabla \left(\nabla\cdot \vec E_1\right)
 =0. \label{e12}
\ee
Here $L_n=(n_0/n_0')$ is the characteristic scale-length for the equilibrium density gradient,  $n_0'=dn_0/dx$, and $L_n=-L_{\sss T}$.

The $y$-component of Eq.~(\ref{e12}) yields one  TEM  wave $\omega^2= \omega_{pa}^2 + \omega_{pb}^2 + k^2 c^2$ which is decoupled from the rest.
The $z$ and $x$ components are coupled and they yield
\[
\left[ - \omega^2 + \omega_{pa}^2 + \omega_{pb}^2 + k^2 \left(\frac{\omega_{pa}^2 \vta^2}{\omega_a^2} + \frac{\omega_{pb}^2 \vtb^2}{\omega_b^2}\right)\right]E_{1z}
\]
\be
-\frac{i k }{L_n}
\left(\frac{\omega_{pa}^2 \vta^2}{\omega_a^2}
+ \frac{\omega_{pb}^2 \vtb^2}{\omega_b^2}\right) E_{1x}=0,\label{e13}
\ee
\[
\left( - \omega^2 +  \omega_{pa}^2 + \omega_{pb}^2 + k^2 c^2\right)E_{1x}
\]
\be
+\frac{i k^3 }{L_n}
\left(\frac{\omega_{pa}^2\vta^4}{\omega_a^4} + \frac{\omega_{pb}^2\vtb^4}{\omega_b^4}\right) E_{1z}=0.\label{e14}
\ee
Eqs.~(\ref{e13}, \ref{e14}) describe coupled longitudinal  $E_{1z}$ and transverse (electromagnetic) $E_{1x}$ perturbations.

Obviously, the coupling described by  Eqs.~(\ref{e13}, \ref{e14}) vanishes in the absence of inhomogeneity $L_n\rightarrow \infty$. In this limit Eq.~(\ref{e14}) yields yet another EM transverse  wave with the electric field in  the $x$-direction, while in e-i plasmas Eq.~(\ref{e13}) describes the usual IA, IP, and Langmuir modes.

 As Fig.~\ref{mech} suggests, the  high-frequency longitudinal electric field (produced by the transverse EM wave) will cause simultaneous high-frequency density oscillations in the $z$-direction. In the  presence of an additional independent longitudinal (ion-acoustic, or electron Langmuir, or ion-plasma)  mode propagating  in the $z$-direction, obviously there may  be coupling of these two longitudinal displacements of completely different origin, and this is what  Eqs.~(\ref{e13}, \ref{e14})  describe, but this is not all. In fact, as it  will be shown below, there appears an extra oscillatory longitudinal mode, which is partly backward and very low-frequency.

 From all this it is self-evident  that $\vec E_{trans}$  does not have to be polarized strictly in the $(x, z)$-plain. To have the described high-frequency longitudinal motion caused by the transverse wave, it is enough that the electric field vector of the transverse wave  has a component in the $x$-direction regardless how small, as we stressed earlier.

So in the presence of inhomogeneity,  the dispersion equation obtained from Eqs.~(\ref{e13}, \ref{e14}) reads:
\[
\omega^2 \left(1-\frac{\omega_{pa}^2}{\omega_a^2} - \frac{\omega_{pb}^2}{\omega_b^2}\right) \left(c^2 k^2 + \omega_{pa}^2 + \omega_{pb}^2 - \omega^2\right)
\]
\be
+\frac{k^4}{L_n^2}  \left(\frac{\omega_{pa}^2\vta^2}{\omega_a^2} + \frac{\omega_{pb}^2\vtb^2}{\omega_b^2}\right) \left(\frac{\omega_{pa}^2\vta^4}{\omega_a^4} + \frac{\omega_{pb}^2\vtb^4}{\omega_b^4}\right)=0. \label{e17}
\ee
Here, $\omega_{a, b}^2\approx \omega^2- k^2 v_{{\sss T}a,b}^2$. In the absence of inhomogeneity this equation describes the usual TEM, Langmuir, and IA modes. With equilibrium gradients it yields some extra branches, like the ion thermal and more importantly a completely new low frequency mode described below for electron-ion and pair-ion plasmas.

\subsubsection{\label{ss2} Electron-ion plasma}

\begin{figure}[!htb]
   \centering
  \includegraphics[height=6.5cm,bb=17 14 264 217,clip=]{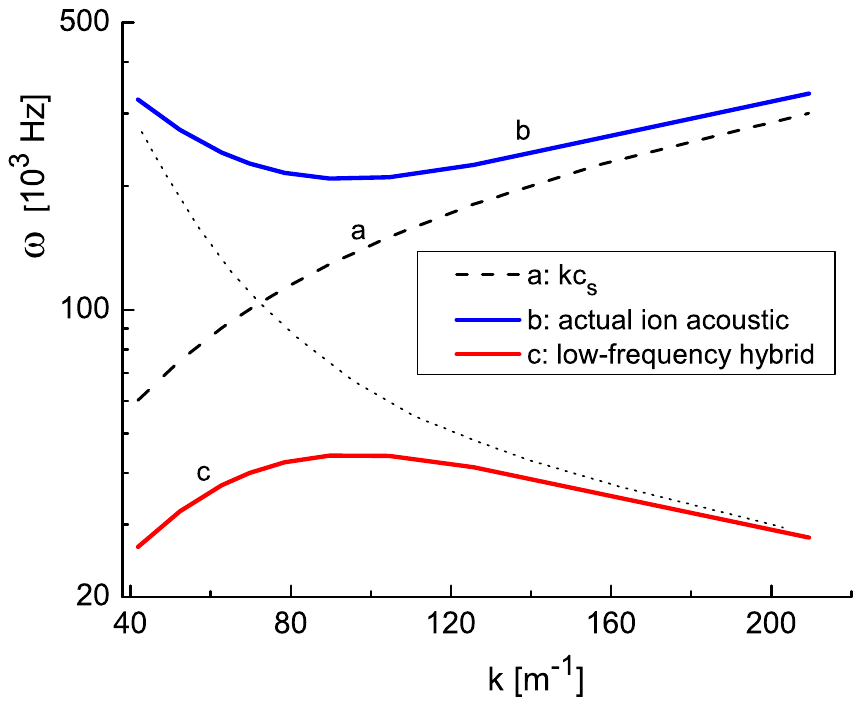}
      \caption{Dispersion equation Eq.~(\ref{e17}) solved for argon-electron plasma. Only ion acoustic and low-frequency hybrid modes are presented.    }   \label{ia}
       \end{figure}

\paragraph{Ion acoustic range.} In electron-ion plasmas we may discuss Eq.~(\ref{e17}) in the IA frequency range, and for simplicity we may assume $T_a\ll T_b$, where $a$ denotes ions $a=i$, and therefore $b=e$. Assuming only that $\omega^2\ll \omega_{pi}^2, \omega_{pe}^2, k^2\vte^2$ (all well-justified for the IA frequency range), this yields the modified IA mode\citep{vaa,v07}
\be
\omega^2= k^2 c_s^2 \left(1+ \frac{m_i}{m_e} \frac{1}{k^2L_n^2} \frac{1}{1+ k^2\lambda_e^2}\right)\frac{1}{1+ k^2\lambda_{de}^2}, \label{e17a}
\ee
 \[
 \lambda_e=\frac{c}{\omega_{pe}}, \quad \lambda_{de}=\frac{\vte}{\omega_{pe}}, \quad c_s^2=\frac{\kappa T_e}{m_i}.
 \]
 The IA mode is partly backward, $\partial \omega/\partial k<0$, due to the second term within brackets in Eq.~(\ref{e17a}), and this is  in the range of small $k$, i.e., for a strong enough inhomogeneity:
 \[
 L_n^2<\frac{m_i}{m_e} \frac{\lambda_e^2}{(1+ k^2\lambda_e^2)^2}.
 \]
 In this $k$-range the IA mode does not go to zero following the usual $k c_s$ line  $a$ [which here is not a straight line due to logarithmic $\omega$-scale]. Instead, the frequency is increased [see Fig.~\ref{ia}] and various consequences of this are discussed in detail in Ref.~4, including the reduced Landau damping. As a result, the IA mode may be expected even in hot-ion plasmas like the solar corona\citep{vaa}. This all is just the consequence of the plasma inhomogeneity.

\paragraph{Sub-IA range: new hybrid mode.} The described backward features of the IA mode are a part of a more global picture which could be partly seen by solving the dispersion equation (\ref{e17}) numerically and analytically. Keeping the ion thermal terms and in the low frequency limit $\omega\ll k \vti$  this yields a new low-frequency hybrid (LFH) mode in an electron-ion plasma with  the frequency which in the given limit reads:
\be
\omega^2\simeq \frac{\lambda_d^2}{L_n^2} \frac{\omega_{pe}^2 + \omega_{pi}^2}{(1+ k^2 \lambda_{ei}^2)(1+ k^2 \lambda_d^2)},
\label{e19a}
\ee
\[
\lambda_d^2=\frac{\lambda_{di}^2 \lambda_{de}^2}{\lambda_{di}^2 + \lambda_{de}^2}, \quad \lambda_{ei}^2=\frac{c^2}{\omega_{pe}^2+ \omega_{pi}^2}.
\]
This mode is  backward for large wavenumber $k$,  it has both electromagnetic and electrostatic features (the electric field of the mode includes  two components, one parallel to the wave-number, and another transverse component  in the direction of the density gradient), and it appears only in an inhomogeneous plasma  due to linear coupling between TEM  and LES waves. Because the frequency increases for decreased $k$, in the small-$k$ range the frequency line would cross the IA dispersion line $kc_s$ [see the dotted line in Fig.~\ref{ia}], but instead of this the two modes exchange identities in such a way that the IA mode becomes backward in the range above $k c_s$ as described by Eq.~(\ref{e17a}), while the LFH mode becomes a direct mode which goes to zero following the line $k c_s$ for $k\rightarrow 0$.

These all features can be seen in Fig.~\ref{ia} which shows the actual ion acoustic wave (line $b$) in an arbitrary inhomogeneous ($L_n=0.5$ m)  argon-electron plasma with $n_0=10^{16}$ m$^{-3}$, $T_e=10^4$ K, $T_i=2\cdot10^3$ K. This line is nicely described by the approximate analytical expression (\ref{e17a}). The usual IA mode $k c_s$ in a homogenous plasma is presented by line $a$. Clearly, in the small $k$ range it is very different from the actual IA wave in inhomogeneous plasma, line $b$.  The new, gradient-driven low frequency hybrid mode (LFH) is presented by line $c$; its maximum frequency in the graph is about  44 kHz and {\bf it is achieved}  at $\lambda=0.06$ m, i.e., $k=90$ (note that the corresponding ion acoustic wave minimum in the same $k$-range is about 208 kHz). For large $k$ (i.e., in the range where it is backward), the approximate analytic expression (\ref{e19a}) describes the LFH mode rather accurately.

\subsubsection{\label{ss3} Pair-ion plasma}

 In pair-ion plasmas\citep{h1, h2, h4, h5, k, vk1, vk2}  without density gradient Eq.~(\ref{e17}) yields the following dispersion equation:
\be
1=\frac{\omega_p^2}{\omega^2- k^2 \vta^2}   + \frac{\omega_p^2}{\omega^2- k^2 \vtb^2}, \quad \omega_p^2=\frac{e^2 n_0}{\varepsilon_0 m}.
\label{p16}
\ee
Eq.~(\ref{p16}) is discussed in our earlier work.  \cite{v08} It gives a longitudinal electrostatic Langmuir mode, and the ion sound mode in the pair plasma (the latter only on condition  $T_a\neq T_b$).

In the presence of the density/temperature gradients, the  LFH  mode with mixed transverse-longitudinal features can be found in the pair-ion plasma  as well.  In the limit $\omega^2\ll \omega_p^2, \, k^2 \vt^2$, Eq.~(\ref{e17}) reduces to
\be
\omega^2=\frac{\vt^2}{L_n^2} \frac{1}{(1+ k^2 r_d^2/2)(1+ k^2 \lambda_{in}^2/2)}.
\label{p18}
\ee
Here, $\lambda_{in}=c/\omega_p$ is the inertial length for the two species, and $r_d=\vt/\omega_p$ is the plasma Debye radius.
  Eq.~(\ref{p18}) is an approximate solution valid only for relatively large wave-numbers;  in this range of $k$ it is also backward,  it should have similar features as the mode in e-i plasma, in particular in the limit  $k\rightarrow 0$, but  this cannot be checked analytically because the condition $\omega^2\ll  k^2 \vt^2$ becomes violated.

\begin{figure}[!htb]
   \centering
  \includegraphics[height=6.5cm,bb=16 14 270 216,clip=]{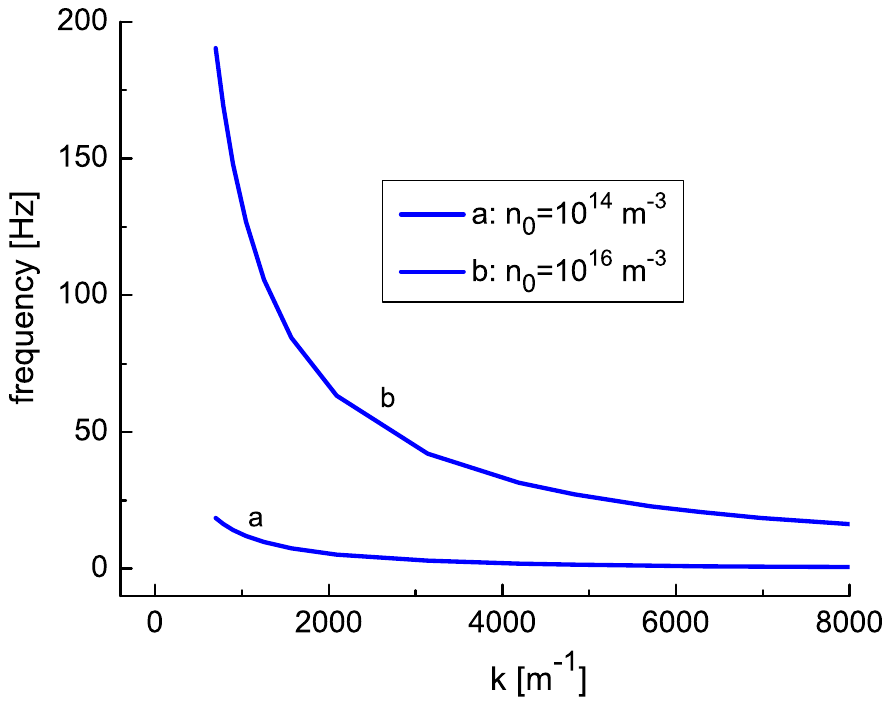}
      \caption{New backward mode in inhomogeneous pair $H^{\pm}$ plasma.}   \label{new}
       \end{figure}

    Eq.~(\ref{e17}) can be  solved numerically for any pair plasma (electron-positron, hydrogen pair plasma H$^\pm$, or fullerene pair plasma). In cylindric configuration in pair-plasma experiments the radial scale was very small, so now we choose $L_n=0.03$ m.  In Fig.~\ref{new} we present only the new hybrid mode  for an arbitrary pair-proton plasma with the temperature  $T_a=T_b=5\cdot 10^3$ K, and  for two densities   $n_0=10^{14}$ m$^{-3}$, and $n_0=10^{16}$ m$^{-3}$.
The approximate solution given by Eq.~(\ref{p18}) is practically identical to the numerical one presented in Fig.~\ref{new} which is obtained from  Eq.~(\ref{e17}).

For the given $k$-range the thermal mode $k \vt$ is with the frequency in the range above  $ 10^6$ Hz and the plasma frequency is $ \sim 10^7$ Hz, so  the new mode presented in Fig.~\ref{new} is far  separated from these frequencies, and it is truly a different branch of oscillations. Note that  $r_d\ll \lambda\ll \lambda_{in}$ so the backward features are mainly due to $k^2 \lambda_{in}^2$ term in Eq.~(\ref{p18}).

For a larger laboratory configuration the mode may be observable by naked eyes. Taking for example $L_n=0.2$ m and $\lambda=0.01$ m for the same temperature as above and $n_0=10^{14}$ m$^{-3}$ yields $\omega=3$ Hz.

The same behavior can be shown by taking fullerene pair plasma, but frequencies in that case are far below Hz.

Taking  parameters which would correspond to experiments \citep{h1, h2, h4, h5} is not appropriate  because of the following two major problems: particle gyro radius {\bf becomes}  comparable with the plasma column radius, and  the density scale length $L_n$ (in radial direction) {\bf is}  much shorter than the parallel wavelength. In such a case an eigen-mode analysis is needed\citep{vpp} and the local approximation is not applicable, so this is  avoided here in order to show that the new mode and its backward features are not the result of the finite geometry and boundary effects.

\section{\label{s4} Plasma in magnetic field}

 In the laboratory environment, instead of the equilibrium condition (\ref{e2}) we may have a more realistic situation of a   plasma  confined by an external magnetic field.  Hence, we shall assume the presence of a background magnetic field directed along the  $z$-axis,  and with perturbations propagating in the same direction. In such a case the equilibrium implies the presence of the diamagnetic drift speed
 \be
 \vec v_{j0}=\frac{\vec e_z\times \nabla p_{j0} }{q_j n_0 B_0}, \quad p_{j0}= \kappa T_{j0} n_{j0}. \label{m1}
 \ee
  The equilibrium may  obviously be satisfied even for a homogeneous temperature and this is a frequent situation in lab-plasmas.  Also, in pair-ion experiments  no temperature gradient is mentioned, so we can omit it in (\ref{m1}) and keep only the density gradient. In such a case, and for other parameters as  above for electron ion plasma we have $v_{e0}=1.7/B_0$ m/s, and for the pair-ion plasma parameters $v_{j0}=14.3/B_0$ m/s,  so an eventual magnetic field  shear inhomogeneity caused by this diamagnetic current is negligible.  On the other hand,  eventual equilibrium magnetic field gradient, which may in principle appear in the equilibrium condition $\nabla [p_0 + B_0^2/(2 \mu_0)]=0$,  can also be made negligible  if plasma-$\beta$ is  small,\citep{krall,vmnras} and this is the case for the parameters used so far in the text.
  Note also that $\vec v_{j0}$ will not contribute to the previously used continuity equation through the term $n_1\nabla\cdot \vec v_{j0}$  because $\nabla\cdot \vec v_{j0}\equiv 0$.

Another issue are the
terms $\vec v_1\times \vec B_0\equiv \vec v_{\bot 1}\times \vec B_0$  and $\vec v_0\times \vec B_1$ that should appear in the momentum equation (\ref{e1})  in the presence of the magnetic field. Regarding their role with respect to the TEM wave, the first can be made negligible because  the Lorentz force due to this term acts on the particle within time interval that is far shorter than the gyro-rotation time (particle changes direction within very short time intervals corresponding to the TEM wave). Note that for the pair-ion case  discussed in the previous section  and assuming $B_0=0.3$ T (like in the experiment) for the two densities we have $\omega_p/\Omega_i\approx 2,\, 22$,  where $\Omega_i=q_i B_0/m_i$. The frequency of the electromagnetic wave is in fact much higher than $\omega_p\simeq 10^6$ Hz; for the smallest $k$ in Fig.~\ref{new} it is of the order of $10^{11}$ Hz.  So it would be  surely justified to omit $\vec v_1\times \vec B_0$ term {\bf even for the parameters used so far for plasma without the magnetic field}. The second term $\vec v_0\times \vec B_1$ is even smaller.

For  the electron-ion plasma the Lorentz force {\bf due to TEM wave} can make difference for electrons only, but it can also be negligible if the magnetic field is not too strong. For example, for the parameters used in Fig.~\ref{ia} the physics  related to the TEM mode alone will not change if the introduced magnetic field is kept below 0.01 T, but the field can be allowed to be much stronger if the number density is assumed higher.

However, in the case of the LFH mode studied in the present text, there appears a slow transverse particle dynamics associated with this mode as well, and this can further be affected by the background field. For this motion there are several possibilities which can be discussed separately.

 1) Gyro-effects will not appear for unmagnetized particles, which corresponds to the frequency limit:
\be
\Omega_j^2< \omega^2\ll \omega_{pi}^2, k^2\vti^2. \label{fr}
\ee
 In pair plasmas (e.g., electron-positron, $H^{\pm}$)  the condition (\ref{fr}) is satisfied in a weak field $B_0^{-5}$ T, and for $T_0=10^5$ K, $n_0=10^{18}$ m${-3}$, $L_n=0.01$ m.
Hence, in this case  the magnetic field effects can  be omitted  from the terms describing the perpendicular perturbed speed,  and the analysis can  be done similar as before with neglected temperature gradient only. The wave Eq.~(\ref{e11}) now becomes
\[
 c^2k^2 \vec E_1 - c^2 \vec k (\vec k\cdot\vec E_1) - \omega^2 \vec E_1+  \left(\omega_{pa}^2 + \omega_{pb}^2\right) \vec E_1
 \]
\[
-  \left(\frac{\omega_{pa}^2\vta^2}{\omega_a^2} +\frac{\omega_{pb}^2\vtb^2}{\omega_b^2}\right) \left[\frac{n_0'}{n_0} \nabla E_{1x}
+   \nabla(\nabla\cdot\vec E_1) \right.
\]
\be
\left.
+ (\nabla\cdot\vec E_1)\frac{\nabla n_0}{n_0}\right]=0, \quad \omega_j^2=\omega^2-k^2\vtj^2.\label{e19}
\ee
The $y$-component yields again a separate TEM mode and from the other two components
we obtain the dispersion equation
\[
\omega^2 \left(1-\frac{\omega_{pa}^2}{\omega_a^2} - \frac{\omega_{pb}^2}{\omega_b^2}\right) \left(c^2 k^2 + \omega_{pa}^2 + \omega_{pb}^2 - \omega^2\right)
\]
\be
-\frac{k^2 }{L_n^2}  \left(\frac{\omega_{pa}^2\vta^2}{\omega_a^2} + \frac{\omega_{pb}^2\vtb^2}{\omega_b^2}\right)^2=0. \label{e20}
\ee
In the pair plasma, for  $\vta=\vtb=\vt$,   dispersion equation (\ref{e20}) yields  the same expression (\ref{p18}) for the frequency.

2) For electron-ion plasma and for parameters and geometry used in Fig.~\ref{ia} and assuming $B_0=0.01$ T or smaller, the frequency of the LFH mode (line $c$) in the given $k$-range is always above the ion gyro-frequency. So the frequency range (\ref{fr}) applies for ions and their dynamics would remain the same.
  But electrons are magnetized for the same parameters (their gyro-frequency is $10^9$ Hz) and their dynamics would become much more complex. Within some reasonable approximations their perpendicular speed can be written as:
  \be
\vec v_{e\bot 1}\approx -\frac{1}{B_0} \vec e_z\times \vec E_{1z} - \frac{\vte^2}{\Omega_e} \vec e_z\times \frac{\nabla_{\bot} n_{e1}}{n_0}
-\frac{1}{\Omega_e B_0}\frac{\partial E_{1\bot}}{\partial t}. \label{a2}
\ee
 Here, only the last term (the polarization drift) describes the motion in the direction of the TEM wave electric field, and only this electron motion can contribute to the appearance of the longitudinal electric field (see Fig.~\ref{mech}), but it is typically negligible. The other two terms describe electron drift motion in the $y$-direction. Electron parallel dynamics (along the magnetic field) will be the same as before.
 From Fig.~\ref{mech} it is clear that the longitudinal electric field will appear whenever any of the two species move along the density gradient. In the present case this will be mainly due to ion direct motion in  the TEM wave field. So some sort of the LFH mode is expected to develop again, but ions are less mobile and the mode may be considerably modified.

  But in a different geometry, with the TEM wave electric field making an angle $\theta$  with respect to the density gradient, the electron drift speed $\vec v_d$ will have one component $v_d \sin(\theta)$ along the density gradient. Electrons will then considerably contribute to the longitudinal electric field, very much similar to what happens in Fig.~\ref{mech}.

3) Finally, if both species are magnetized $\omega^2< \Omega_j^2$, only drift motion can develop in direction perpendicular to the magnetic field. From Fig.~\ref{mech} we have learned that, to have the effects studied here, there must be displacement (of any kind) of particles (of any species) along the density gradient, and obviously this can be either a direct motion (as presented in the figure) or a drift in the presence of the magnetic field. However, in the present case, the  drift of particles  along the density gradient will develop  only if the electric field of the incident TEM wave is not strictly  in the same direction. So some sort of LFH mode is expected again but geometry should be assumed different from the cases studied above.

In both cases 2) and 3), derivations are lengthy, particle dynamics in two perpendicular directions become coupled,  and several additional modes appear (like electron and ion cyclotron, lower and upper hybrid, etc.). Yet  no essential new physics is expected to emerge and such derivations will be omitted here.

\section{\label{c} Summary and conclusions}

The mechanism of coupling between TEM and LES waves propagating in inhomogeneous plasmas, discovered in Ref.~4, is shown here to contain some crucial extra physics. A completely new mode is shown to exist, which was not noticed in Ref.~4. This new low frequency hybrid (LFH) mode is the result of coupling between transverse and longitudinal electric fields of electromagnetic and electrostatic waves in the presence of density gradient.  So the backward features of the ion acoustic mode, discussed in Ref.~4, are a a part of a more profound and complex phenomenon: for relatively small wave-numbers the backward LFH mode exchanges identity with the IA wave, the latter becomes backward for $k\rightarrow 0$ and eventually gets some cut-off or its frequency continues to grow in this $k$-range, while the LFH mode goes  towards $\omega=0$ following the usual IA line $k c_s$ in the same $k$-limit. These features are partly seen in Fig.~\ref{ia}.  The properties of the LFH mode do not necessarily change in the presence of the magnetic field because transverse motion of particles (with respect to the magnetic field vector) is essentially due to the TEM wave which is of very high-frequency so that  gyro-motion of particles is usually negligible. In the text, the features of the LFH mode are presented for two possible equilibria and for both electron-ion and pair plasmas.

The presented TEM-LES coupling is linear, and it happens only in the presence of a density gradient. However, for some  nonlinear phenomena the LFH mode may take over the role of the usual IA wave. The obvious possibilities are the following: electron decay instability (LHF mode interacting with two Langmuir waves propagating in opposite directions), parametric backscattering (LFH mode interacting with two light waves  propagating in opposite directions), and parametric decay instability (incident light wave interacting with  Langmuir wave and LFH mode moving in opposite direction).

 The two different equilibria discussed in Secs.~\ref{s3}, \ref{s4} have effect on the dispersion equation in general [see the coupling term containing $L_n$ in dispersion equations (\ref{e17}, \ref{e20})], although in the case of the  LFH mode this effect vanishes in approximate expressions. This approximate  absence of the effect on the LFH mode may be explained in the following way.  As pointed out earlier, in the case $\nabla p_{j0}=0$ [no magnetic field, the condition (\ref{e2})] the last two terms in equation (\ref{e11}) vanish, and the remaining {\em temperature gradient} terms can be traced in Eqs.~(\ref{e12}, \ref{e17}) through the terms which contain $\vtj^4$. On the other hand, in the presence of magnetic field, the temperature gradient is assumed absent because it is not essential or required for the equilibrium, so these terms are set to zero in Eq.~(\ref{e11}). Now the remaining {\em density gradient} terms yield the term $\alpha=\omega_{pa}^2\vta^2/\omega_a^2 +\omega_{pb}^2\vtb^2/\omega_b^2$ in Eq.~(\ref{e19}) that  appears with both $\vec E$-field components $E_{1x}, E_{1z}$, which are however coupled and this then yields $\alpha^2$ in the dispersion equation (\ref{e20}). But in the LFH mode limit all the terms $\omega_j^2=\omega^2-k^2\vtj^2$  reduce to $-k^2\vtj^2$, and then there is an obvious cancelation of the remaining thermal terms in the coupling term with $L_n$, so  these equilibrium differences vanish. Though strictly speaking this is so only for relatively large values of $k$, as can be deduced from Fig.~\ref{ia} where neglecting $\omega^2$ in the small $k$ limit is not justified any longer.

\nocite{*}
\bibliography{aipsamp}

\vfill
\eject

\end{document}